\documentstyle[12pt]{article}
\begin{document}

\def\lsim{\lower .5ex\hbox{$\buildrel < \over {\sim}$}}
\def\gsim{\lower .5ex\hbox{$\buildrel > \over {\sim}$}}

\tolerance=10000 
\raggedbottom  
\baselineskip=12pt  

\centerline{\bf Liquid metallic hydrogen and the structure of brown dwarfs and giant planets}

\vskip 0.3cm
\centerline{W.B. Hubbard, T. Guillot, J.I. Lunine}
\centerline{\it Lunar and Planetary Laboratory, University of Arizona, Tucson, AZ 85721}
\centerline{hubbard@lpl.arizona.edu}
\vskip 0.2cm

\centerline{A. Burrows}
\centerline{\it Departments of Physics and Astronomy, University of Arizona, Tucson, AZ 85721}
\vskip 0.2cm

\centerline{D. Saumon}
\centerline{\it Department of Physics and Astronomy, Vanderbilt University, Nashville, TN 37235}
\vskip 0.2cm

\centerline{M.S. Marley}
\centerline{\it Department of Astronomy, New Mexico State University, Las Cruces, NM 88003}
\vskip 0.2cm

\centerline{R.S. Freedman}
\centerline{\it Sterling Software, NASA Ames Research Center, Moffett Field, CA 94035}
\vskip 0.3cm

\centerline{\bf ABSTRACT}

Electron-degenerate, pressure-ionized hydrogen (usually referred to as
metallic hydrogen) is the principal constituent of brown dwarfs, the
long-sought objects which lie in the mass range between the lowest-mass
stars (about eighty times the mass of Jupiter) and the giant planets.
The thermodynamics and transport properties of metallic hydrogen are
important for understanding the properties of these objects, which,
unlike stars, continually and slowly cool from initial nondegenerate
(gaseous) states.

Within the last year, a brown dwarf (Gliese 229 B) has been detected and
its spectrum observed and analyzed, and several examples of extrasolar
giant planets have been discovered.  The brown dwarf appears to have a
mass of about forty to fifty Jupiter masses and is now too cool to be fusing
hydrogen or deuterium, although we predict that it will have consumed all of its
primordial deuterium.  This paper reviews the current understanding of the
interrelationship between its interior properties and its observed
spectrum, and also discusses the current status of research on the 
structure of giant planets, both in our solar system and elsewhere.
\vskip 0.3cm


\vfil
\eject
\noindent{\bf I. INTRODUCTION}

For the purposes of this paper, we define a giant planet to be a hydrogen-rich
object similar
to Jupiter (mass $M_J=1.9 \times 10^{30}$ g = 0.95 $\times 10^{-3} M_{\odot}$,
where $M_{\odot}$ is the solar mass) or Saturn (mass $M_S=0.3 M_J$), excluding smaller
objects similar to Uranus or Neptune which have only small mass fractions of hydrogen.  We
also define a brown dwarf to be an object with mass smaller than the minimum mass required
for sustained thermonuclear fusion of hydrogen (about 85 $M_J$), but larger than the
minimum mass required for fusion of deuterium (about 13.5 $M_J$).  In fact the latter
distinction is somewhat arbitrary, for giant planets and brown dwarfs constitute a
seamless continuum, with the same interior physics applying to both.  

There has long been a wealth of theoretical studies of the properties of giant planets
and brown dwarfs$^{1,2}$, but prior to 1995, the only objects to which these theories
could be applied were Jupiter and Saturn.  However, the year 1995 marked an
explosive turning point in this situation.  Within an interval of a few months, the first
detections of extrasolar giant planets were reported$^3$, and a bona fide brown dwarf,
Gl229 B, was discovered$^4$.  In rapid succession, high-quality spectral data on the emissions
of the new brown dwarf were obtained, permitting an initial determination of its atmospheric
composition$^5$, and an entry probe carried out
{\it in situ} measurements of the structure and composition of the Jovian atmosphere
to a pressure of 20 bars$^6$.  At the same time, developments in high-pressure technology
have allowed new experimental studies of hydrogen in the megabar pressure range$^7$,
and developments in high-pressure theory have given new information about possible
phase transitions in hydrogen in the relevant pressure range$^8$.

Interpretations of
the new data are dependent upon, and potentially bear information about, the phase diagram and thermodynamic
properties of liquid metallic hydrogen at temperatures $T$ of $\sim 10^4$ to $\sim 10^6$ K, and pressures
$P$ of $\sim 1$ to $\sim 10^6$ Mbar.  In this paper we consider theoretical interpretations of
data for the three main types of objects which relate to the recent observations: 
(a) brown dwarfs (e.g., Gl229 B); (b) Jupiter;
(c) high-temperature
giant planets which orbit at small separations from their primary (e.g., 51 Peg B).
Our primary focus is upon tests of the hydrogen
equations of state and proposed hydrogen phase diagrams.
\vskip 1cm

\noindent{\bf II. HYDROGEN PHASE DIAGRAM}

The relevant parts of the hydrogen phase diagram are shown in Fig. 1.
This diagram is computed assuming pure hydrogen, although, as we shall
discuss below, the non-hydrogen components of brown dwarfs and giant
planets constitute an important diagnostic for this diagram.  The
upper-right-hand corner of the experimentally-accessible regime is
defined by the open circle, which shows the highest temperature and
pressure investigated in recent high-pressure electrical-conductivity
experiments$^7$.  The solid circle shows an experimentally-observed
transition to a metallic state of hydrogen at $P=1.4$ Mbar and $T=3000$ K.
At pressures below the solid circle, electrical conductivity of
hydrogen increases rapidly with pressure, while for pressures
lying between the two circles, the measured electrical conductivity
is essentially constant at 2000 $(\Omega - {\rm cm})^{-1}$, but
still about two orders of magnitude lower than the value predicted
for classical fully-ionized liquid metallic hydrogen at comparable
density and temperature.  In the experimental data, there is no
evidence of a discontinuous change in state as a function of
pressure up to the maximum pressure.  These data$^7$ indicate that
hydrogen dissociates continuously from the molecular to the
metallic phase with no first-order transition.  Furthermore, the
gradual latent heat release associated with the gradual dissociation
causes a cooling effect on interior temperatures similar to the
effect of a radiative zone in Jupiter and cooler brown dwarfs
described below.  A fully self-consistent equation of state
for models of these objects which fully incorporates the new
experimental data is not yet available.

The conductivity measurements bear upon theoretical calculations
of the hydrogen phase diagram at temperatures near $10^4$ K and pressures
$\sim 0.1$ to 1 Mbar.  Figure 1 shows two such caculations, which
have qualitatively similar results, predicting that a phase transition
exists between two liquid phases of hydrogen.  The calculation by
Saumon {\it et al.}$^8$ is based upon a chemical approach which evaluates
the equilibrium between various neutral and ionized forms of H and
H$_2$, while the calculation by Magro {\it et al.}$^8$ is based upon
first-principles evaluation of the quantum partition function of
a system of electrons and protons.  Both calculations predict a critical point
at a temperature slightly above $10^4$ K and a pressure somewhat below
1 Mbar, but neither extend into the range where electrical conductivity
was measured.  In this connection it should be mentioned that shock-compression
experiments up to $T=5200$ K and $P=0.83$ Mbar show no evidence of
a major discontinuity in density, although a small discontinuity
($\lsim 10$\%) could remain undetected$^7$.

Also shown in Fig. 1 is the experimentally-observed boundary between
liquid and solid phases of H$_2$, extrapolated to $P \approx 1.4$ Mbar$^9$,
and the theoretically-calculated boundary between liquid and solid
phases of classical metallic hydrogen$^{10}$, assuming that the transition
occurs at a proton plasma coupling parameter $\Gamma=175$, and making
no correction for electron screening or quantum oscillations of the
protons (both of which will lower the melting temperature).  Vertical
dashed lines show guesses for the possible behavior of the boundary
between the insulating and metallic liquid phases (upper dashed line) and
the insulating and metallic solid phases (lower dashed line).  As we
shall see, neither solid phase is relevant to astrophysical objects.

Finally, at the top of Fig. 1 we show the locus of central points in
brown-dwarf models where 50\% of the luminosity is derived from thermonuclear
conversion of deuterium nuclei to ${}^3$He nuclei; for objects which
achieve central temperatures above this line, all of the deuterium is consumed
during the first $\sim 0.01$ Gyr of the objects' existence$^{11}$.  The
dot marking the end of this locus shows the center of a critical model which
is just able to ignite deuterium fusion.  This occurs under conditions of
intermediate proton (deuteron) coupling with $\Gamma \approx 2$.
\vskip 1cm

\noindent{\bf III. EVOLUTIONARY SEQUENCES}

The region of the hydrogen phase diagram which is probed by evolving
giant planets and brown dwarfs is shown in Fig. 2.  This figure shows,
on the same plane as Fig. 1, calculated $T$ vs. $P$ profiles for the
interior of Jupiter (shorter lines) and the interior of Gl229 B
(longer lines), at various ages.  The profiles are isentropes
calculated for a solar mixture of hydrogen and helium (28 \% helium
by mass), using the Saumon-Chabrier-Van Horn (SCVH) equation of state$^8$.
For objects in this mass range, the interior is maintained in a state
of efficient convection by virtue of the relatively low thermal
conductivity of metallic hydrogen and the relatively high opacity
of molecular and metallic hydrogen (with the exception of a radiative
zone which appears in Jupiter and in cooler brown dwarfs at temperatures
between 1200 and 2900 K)$^{12}$.  As a result the temperature profile remains
close to an isentrope throughout the interior, with the specific
entropy defined by the radiative properties of the optically-thin outer
layers.

The evolution of a giant planet or brown dwarf is calculated by using
a relation of the form $S/N = f(T_{\rm eff},g)$, where $S/N$ is the
specific entropy (entropy per nucleon), $T_{\rm eff}$ is the effective
temperature of the atmosphere, and $g$ is the surface gravity.
Calculation of $f$ involves solution of the coupled equations
of radiative/convective heat transport and hydrostatic equilibrium
for specified $T_{\rm eff}$ and $g$ in the relevant ranges ($T_{\rm eff}=
124$ K and $g=23$ m/s$^2$ for Jupiter, to $T_{\rm eff}=
960$ K and $g=1000$ m/s$^2$ for Gl229 B), and is nontrivial, as atmospheres
in the relevant ranges have extremely nongrey fluxes (for Gl229 B the
predicted fluxes vary by as much as three orders of magnitude over
adjacent wavelength bands), and trace species such as methane, water,
and ammonia have major influences on the fluxes.  Mapping of the
function $f$ over the relevant $T_{\rm eff},g$ range is still in
progress, but initial results$^{11}$ have been obtained for the
evolution of objects in the mass range from $\sim 80$ $M_J$
down to $\sim 1$ $M_J$. 

The first general calculation of the evolution of giant planets and
brown dwarfs was carried out by Black$^{13}$, who fitted power laws to
existing calculations for the evolution of Jupiter and brown dwarfs,
and used the results to predict the behavior of objects in the range
from $\sim 1$ to 15 $M_J$.  Saumon {\it et al.}$^{11}$ improved upon this
calculation by using quantitative theories for the interior thermodynamics
and thermonuclear reaction rates of metallic-hydrogen liquids, and
interpolating in a grid of values for $f$ 
for Jovian and near-Jovian $T_{\rm eff}, g$ in the low end, and
previously-calculated values for brown dwarfs with grey model atmospheres
with $M > 15 M_J$ at the
high end.  Some sample results
from the latter calculation are shown in Fig. 3.  This figure displays
the radii of giant planets as a function of their mass for logarithmically-equal
time intervals from an age $t=10^{-3}$ Gyr to 4.6 Gyr (present age of Jupiter).
As is well known, the compressibility of liquid metallic hydrogen is such
that an approximately constant radius of around $10^5$ km is maintained
throughout the indicated mass and age range.  For fixed age, a minimum
radius (at mass $M=2$ $M_J$) at about 120000 km at $t=10^{-3}$ Gyr is replaced
by a maximum radius at about 78000 km, slightly larger than Jupiter's present
radius, at mass $M=4$ $M_J$ and $t=4.6$ Gyr.

With the discovery of Gl229 B and the availability of a high-quality
spectrum of this object, it has become imperative to generate nongrey
model atmospheres to fit the observations, and to simultaneously use the
atmospheres to determine the best-fit values of $T_{\rm eff}, g$ and thus
$f$.

Figure 4 shows a best-fit model generated by our group
for the atmospheric profile of $T$ vs. $P$ in Gl229 B, along with a
comparison profile for Jupiter$^5$.  At higher pressures (about 40 bar
for Gl229 B and 0.4 bar for Jupiter), the profile merges with an
isentrope (dashed line), the continuation of which is shown in Fig. 2
(lowest profile for each object).

Our updated grid of $f$ values, using the optimized $T(P)$ profiles
which best fit the Gl229 B spectrum, was used to generate an evolutionary
sequence.  A summary of the results is presented in Fig. 5.  It turns out
that at the present stage of fitting Gl229 B's spectrum, the value of $g$
is only poorly constrained, while $T_{\rm eff}$ is better determined.
We find that Gl229 B fits the theory for brown dwarfs extremely well,
and there can be no doubt that this object is a member of the class.
The interior $T(P)$ profiles shown in Fig. 2 correspond to a likely
model inferred from Fig. 5, with a mass of 40 $M_J$ and an age of
1.7 Gyr.   It is noteworthy that the isentrope corresponding to the
present Gl229 B is very close to the Jovian isentrope calculated for
a Jovian age of $\sim 0.1$ Gyr, but of course extends to a central
pressure four orders of magnitude larger.  Because the central pressure
and temperature of Gl229 B greatly exceed the limit for deuterium
fusion, no primordial deuterium should be present in this brown
dwarf's atmosphere.

The interior isentrope corresponding to a jovian planet of $M \sim 1$ $M_J$
heated to $T_{\rm eff} \approx 1250$ K by
close proximity to a primary star (e.g., 51 Peg B),
at an age of 8 Gyr, turns out to be
very similar to a Jovian adiabat at an age of 1 Gyr (see Fig. 2).  A
high-temperature outer radiative zone is created by the insolation
from the primary, but this involves only a small fraction of the
planet's mass and radius$^{14}$.  Because the interior of the planet is at
a much higher entropy at an age of 8 Gyr, the planet is significantly
thermally expanded in radius (by $\sim 20$ \%) compared with the
present Jupiter.  Nevertheless a high-temperature jovian planet similar to
51 Peg B, orbiting at only 0.05 A.U. from its primary, is quite stable.

\vfil
\eject
\noindent{\bf IV. IMPLICATIONS FOR ATMOSPHERIC ABUNDANCES}

Figure 6 shows the phase diagram of Fig. 1 combined with the
evolutionary trajectories of Fig. 2.  The principal implications
of this figure are as follows.  If a high-temperature first-order phase
transition exists between two liquid phases of hydrogen (the
so-called plasma phase transition, or PPT), then according to the
Gibbs phase rule, in thermodynamic equilibrium the concentrations of
nonhydrogen species such as noble gases, water, methane, etc., cannot
be continuous across the boundary.  This mechanism for affecting the
atmospheric abundances of brown dwarfs and giant planets is separate,
and possibly in addition to,
the proposed limited solubility of such species in metallic
hydrogen$^2$.  As Figure 6 indicates, both Jupiter and brown dwarfs
such as Gl229 B will evolve in such a way as to cross the PPT at
later stages in their evolution.  However, it is interesting to
note that at the age that we derive for Gl229 B, it would cross the
PPT calculated by Saumon {\it et al.} but would just miss the
critical point of the PPT calculated by Magro {\it et al.}$^8$

We do not know the effect of the PPT on the partitioning
of nonhydrogen species, for no one has carried out the requisite
calculations.  Moreover, an abundant impurity (helium) will
significantly shift the location of the PPT as well.
In principle one could test for partitioning across the PPT
by comparing the atmospheric abundances of objects such as
Jupiter and Gl229 B.  Water is prominent in the observed
spectrum of Gl229 B, has recently been measured in the
deep Jovian atmosphere by an entry probe$^8$, and should if it has
solar abundance be present at a number fraction of about
$1.5 \times 10^{-3}$ relative to H$_2$.  The synthetic spectrum
of Gl229 B generated by our group, under the assumption of
solar abundance of H$_2$O in the atmosphere, is a good
if not perfect fit to the data$^5$.  Water opacity dominates
many bands of the Gl229 B spectrum, superimposed on a general
background opacity provided by pressure-induced dipole transitions
of H$_2$.  A few bands are indicative of methane and ammonia opacity,
which are also presumed to be present in solar abundance.  While
a quantitative determination of abundances in Gl229 B remains to
be done, at this stage it appears to be safe to say that
its atmosphere has a solar abundance of C, N, and O.

As in Gl229 B, the C, N, and O in Jupiter's observable atmosphere are
primarily in the form of CH$_4$, NH$_3$, and H$_2$O.  The first two
are readily measurable in the Jovian spectrum, and are found to be
enhanced by about a factor of two over solar composition$^6$.  The {\it in situ}
measurements by the Galileo entry probe are more uncertain because they
are still under analysis; Jovian CH$_4$ is found to be present
at 2.9 times solar concentration.  However, the presently-available
interpretation of the probe data suggests that Jupiter's atmospheric H$_2$O
is depleted by about a factor 5 relative to solar$^6$.

There is a known effect, quite independent of the PPT, which could affect
Jupiter's atmospheric water abundance.  In Fig. 4 we have indicated the condensation
line for H$_2$O in a solar-composition atmosphere.  Above this line,
the H$_2$O is in full concentration in the ambient gas, but below this
line, the formation of a condensed phase of H$_2$O will reduce the
gas-phase concentration in accordance with vapor-solid (or liquid) equilibrium.
Note that the atmosphere of Gl229 B is everywhere so warm that its
H$_2$O cannot be affected by this process.
It was expected that by the time the Galileo probe reached the 20 bar level, the
H$_2$O would be fully in solution in the background gas, so a possible
strong depletion at the deepest levels probed would be puzzling.  To explain
the seeming inconsistency with solar composition, Niemann {\it et al.} considered
various alternative scenarios, including extreme downdrafts which might
deplete water in the region entered by the probe$^6$.

A scenario not considered by Niemann {\it et al.} is possible partitioning
by a PPT.  Jupiter's interior isentrope is everywhere much colder than
that of Gl229 B.  Not only does Jupiter's isentrope lie within the
field of H$_2$O condensation, it also lies deep within the PPT of both
current theoretical models for liquid metallic hydrogen (strictly
speaking, the ``isentrope'' would have continuous $T$ and $P$ across
the PPT, and discontinuous entropy and composition).

\vskip 1cm

\noindent{\bf V. CONCLUSIONS}

With the availability of detailed data for the properties and atmospheric
abundances of brown dwarfs and jovian planets, it is becoming possible to
probe the properties of the phase diagram of hydrogen in the vicinity of
the region where dense molecular hydrogen liquid transforms into
metallic hydrogen liquid.  Earlier work$^2$ has shown that
in the metallic-hydrogen liquid, various impurities such as helium
may have limited solubility, and that this may affect atmospheric
abundances.  Here we point out that, quite apart from possible precipitation
within the liquid metallic phase, a first-order phase transition
to the metallic phase will affect relative abundances of species regardless
of their overall concentration.  We are now able to study relative
abundances in an object (Gl229 B) which almost certainly lies above or very near
any PPT critical point, and objects (Jupiter and Saturn) which lie well below
that critical point.  Determination of the existence of
a PPT in hydrogen has therefore become a first-order priority for the
interpretation of new astrophysical data on giant planets and brown dwarfs.
If it can be definitively shown to exist through experiment and theory,
then determination of the partitioning coefficients for various elements
in the two phases will be essential for understanding the interior
structure of giant planets and brown dwarfs.

\vskip 1cm

\noindent{\bf ACKNOWLEDGMENTS}

This work was supported by National Aeronautics and Space
Administration grants NAGW-1555 and NAGW-2817, by
National Science Foundation grant AST93-18970, and by a European
Space Agency fellowship to T.G. and a
Hubble Fellowship to D.S.

\vfil
\eject
\noindent{\bf REFERENCES}
\vskip 0.4cm

{${}^1$} M.S. Kafatos, R.S. Harrington, S.P. Maran,
{\it Astrophysics of Brown Dwarfs} (Cambridge Univ. Press, Cambridge, 1986).
\vskip 0.3cm

{${}^2$} D.J. Stevenson, E.E. Salpeter, Astrophys. J. Suppl.
{\bf 35}, 221 (1977); ibid, 239.
\vskip 0.3cm

{${}^3$} M. Mayor, D. Queloz, Nature {\bf 378}, 355 (1995);
G.W. Marcy, R.P. Butler, Astrophys. J. Lett. {\bf 464}, L147.
\vskip 0.3cm

{${}^4$} T. Nakajima, B.R. Oppenheimer, S.R. Kulkarni, D.A. Golimowski,
K. Matthews, S.T. Durrance, Nature {\bf 378}, 463 (1995);
B.R. Oppenheimer, S.R. Kulkarni, K. Matthews, T. Nakajima, Science
{\bf 270}, 1478 (1995).
\vskip 0.3cm

{${}^5$} M.S. Marley, D. Saumon, T. Guillot, R.S. Freedman,
W.B. Hubbard, A. Burrows, J.I. Lunine, Science {\bf 272}, 1919 (1996).
\vskip 0.3cm

{${}^6$} H.B. Niemann, S.K. Atreya, G.R. Carignan, T.M. Donahue,
J.A. Haberman, D.N. Harpold, R.E. Hartle, D.M. Hunten, W.T Kasprzak,
P.R. Mahaffy, T.C. Owen, N.W. Spencer, Science {\bf 272}, 846 (1996);
U. von Zahn, D.M. Hunten, Science {\bf 272}, 849 (1996).
\vskip 0.3cm

{${}^7$} S.T. Weir, T. Mitchell, W.J. Nellis, Phys. Rev. Lett.
{\bf 76}, 1860 (1996); W.J. Nellis, M. Ross, N.C. Holmes, Science
{\bf 269}, 1249 (1995). 
\vskip 0.3cm

{${}^8$} D. Saumon, G. Chabrier, H.M. Van Horn, Astrophys.
J. Suppl. {\bf 99}, 713 (1995); W.R. Magro, D.M. Ceperley,
C. Pierleoni, B. Bernu, Phys. Rev. Lett. {\bf 76}, 1240 (1996).
\vskip 0.3cm

{${}^9$} I.F. Silvera, in {\it Simple Molecular Systems
at Very High Density}, edited by A. Polian, P. Loubeyre, and
N. Boccara (Plenum, New York, 1989), p. 33.
\vskip 0.3cm

{${}^{10}$} M.D. Jones, D.M. Ceperley, Phys. Rev. Lett.
{\bf 76}, 4572 (1996).
\vskip 0.3cm

{${}^{11}$} D. Saumon, W.B. Hubbard, A. Burrows,
T. Guillot, J.I. Lunine, G. Chabrier, Astrophys. J. {\bf 460},
993 (1996).
\vskip 0.3cm

{${}^{12}$} T. Guillot, G. Chabrier, D. Gautier, P. Morel,
Astrophys. J. {\bf 450}, 463 (1995).
\vskip 0.3cm

{${}^{13}$} D.C. Black, Icarus {\bf 43}, 293 (1980).
\vskip 0.3cm

{${}^{14}$} T. Guillot, A. Burrows, W.B. Hubbard, J.I. Lunine,
D. Saumon, Astrophys. J. Lett. {\bf 459}, L35 (1996).
\vfil
\eject

\noindent{\bf FIGURE CAPTIONS}
\vskip 0.5cm
{\bf Fig. 1 --} The phase diagram of hydrogen at high pressures
and temperatures.  Alternate plasma phase transition lines calculated
by Magro {\it et al.} (QMC stands for ``quantum Monte Carlo''), and
by Saumon {\it et al.} (SCVH) are shown; both terminate in a critical
point$^8$.  ``D burns'' is not a phase transition but rather
the locus of marginal deuterium thermonuclear fusion under conditions
of intermediate plasma coupling.

\vskip 0.3cm
{\bf Fig. 2 --} Interior temperature-pressure profiles for Jupiter
and Gl229 B; the age of the object in Gyr is given at the end of
each profile.  An open box marks the central conditions at each age.
A solid triangle shows central conditions for a jovian-planet
model of the high-temperature extrasolar planet 51 Peg B.

\vskip 0.3cm
{\bf Fig. 3 --} Radius vs. mass at various ages (equally spaced
logarithmically) for jovian planets.  Present values for Jupiter
(J) and Saturn (S) are shown, along with values for the
non-hydrogenic giant planets Uranus (U) and
Neptune (N).  The observed radii of Jupiter and Saturn are slightly
lower than values computed for solar composition because these
planets have interiors enriched in elements heavier than hydrogen and
helium.

\vskip 0.3cm
{\bf Fig. 4 --} Atmospheric temperature profiles for Gl229 B
and Jupiter (solid lines).  At higher pressures, these profiles
merge with isentropes (short-dashed lines).  The long-dashed
curve shows the locus of points below which water in a solar-composition
atmosphere is in equilibrium with a condensed (liquid or solid) phase.

\vskip 0.3cm
{\bf Fig. 5 --} A plot of surface gravity $g$ vs. effective
temperature $T_{\rm eff}$ for brown dwarf models.  Constraints on
effective temperature and surface gravity of Gl229 B are shown
as a shaded central area.  Contours of constant mass (in $M_J$),
radius (in km), and age (in Gyr) are superimposed.

\vskip 0.3cm
{\bf Fig. 6 --} A superimposition of Figs. 1 and 2, showing
positions of Jupiter and Gl229 B evolutionary models with respect
to the hydrogen phase diagram.

\vfil
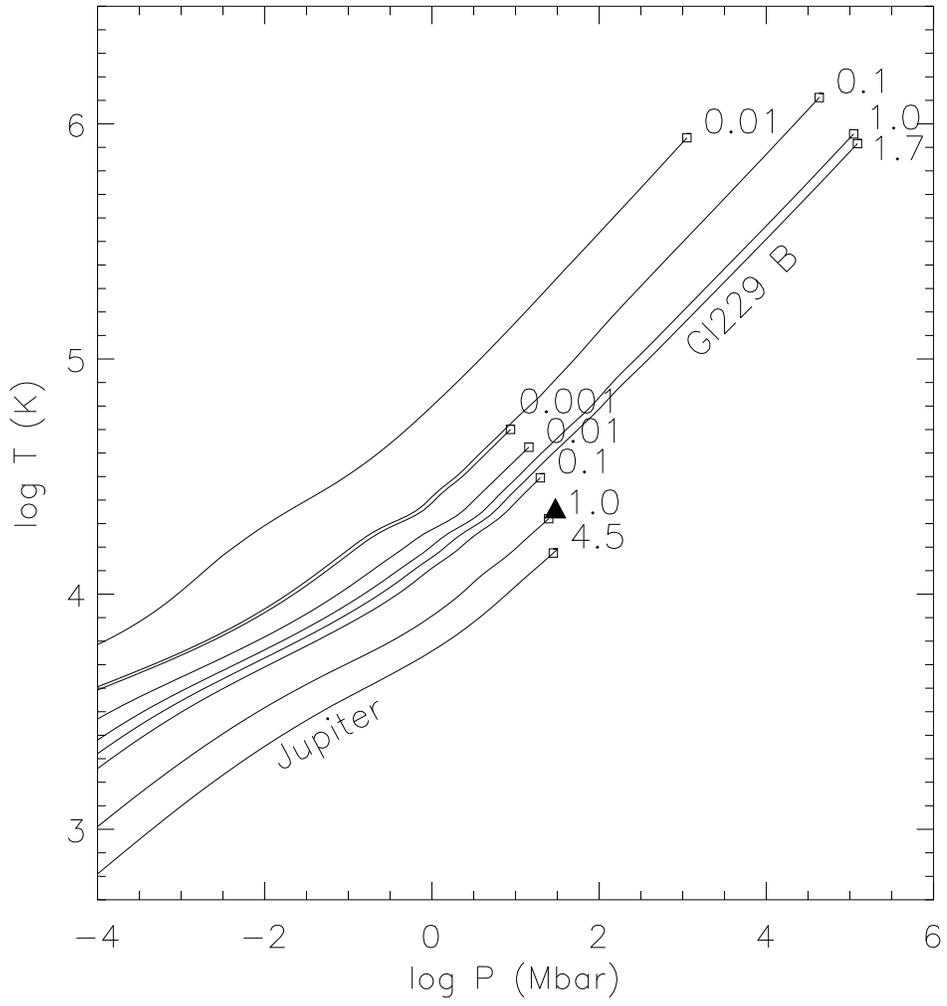
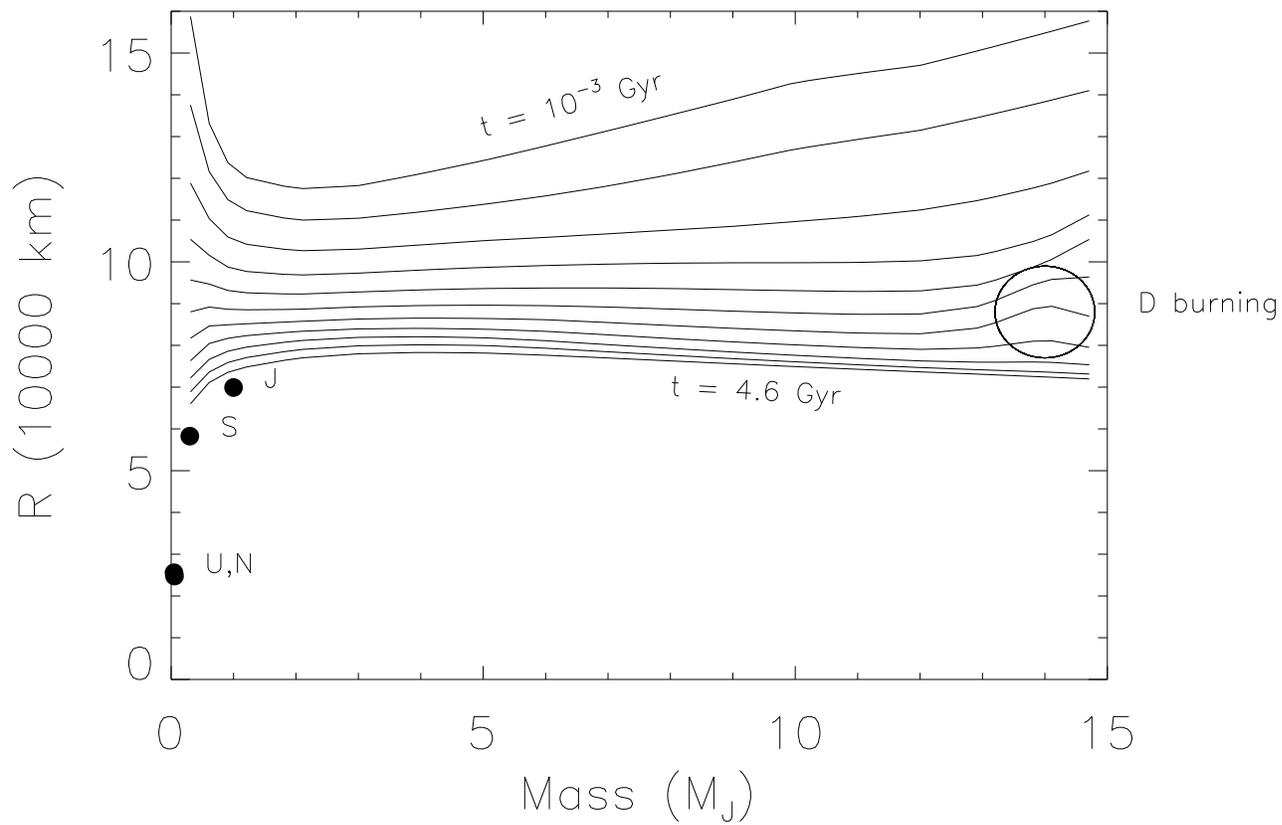
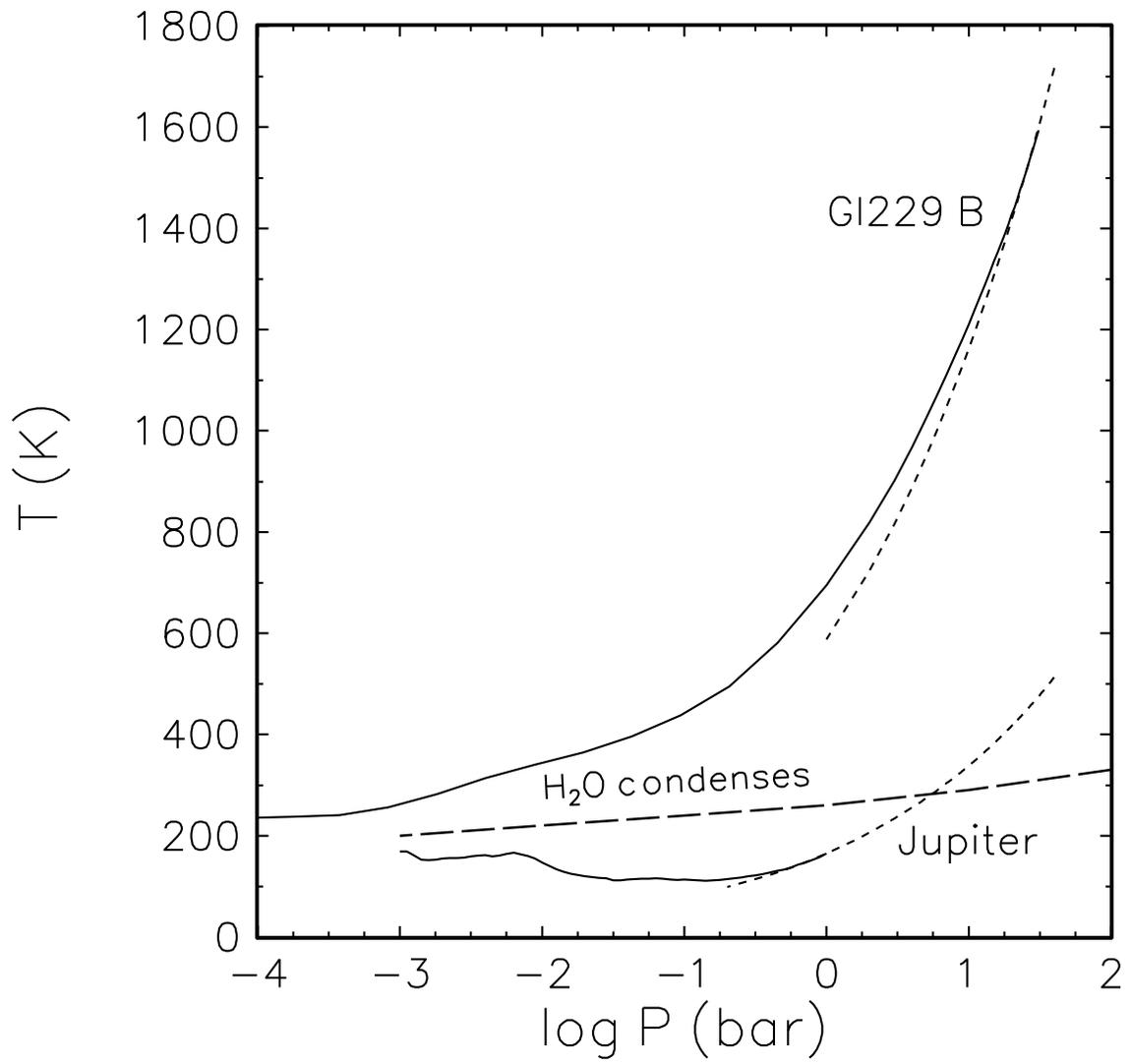
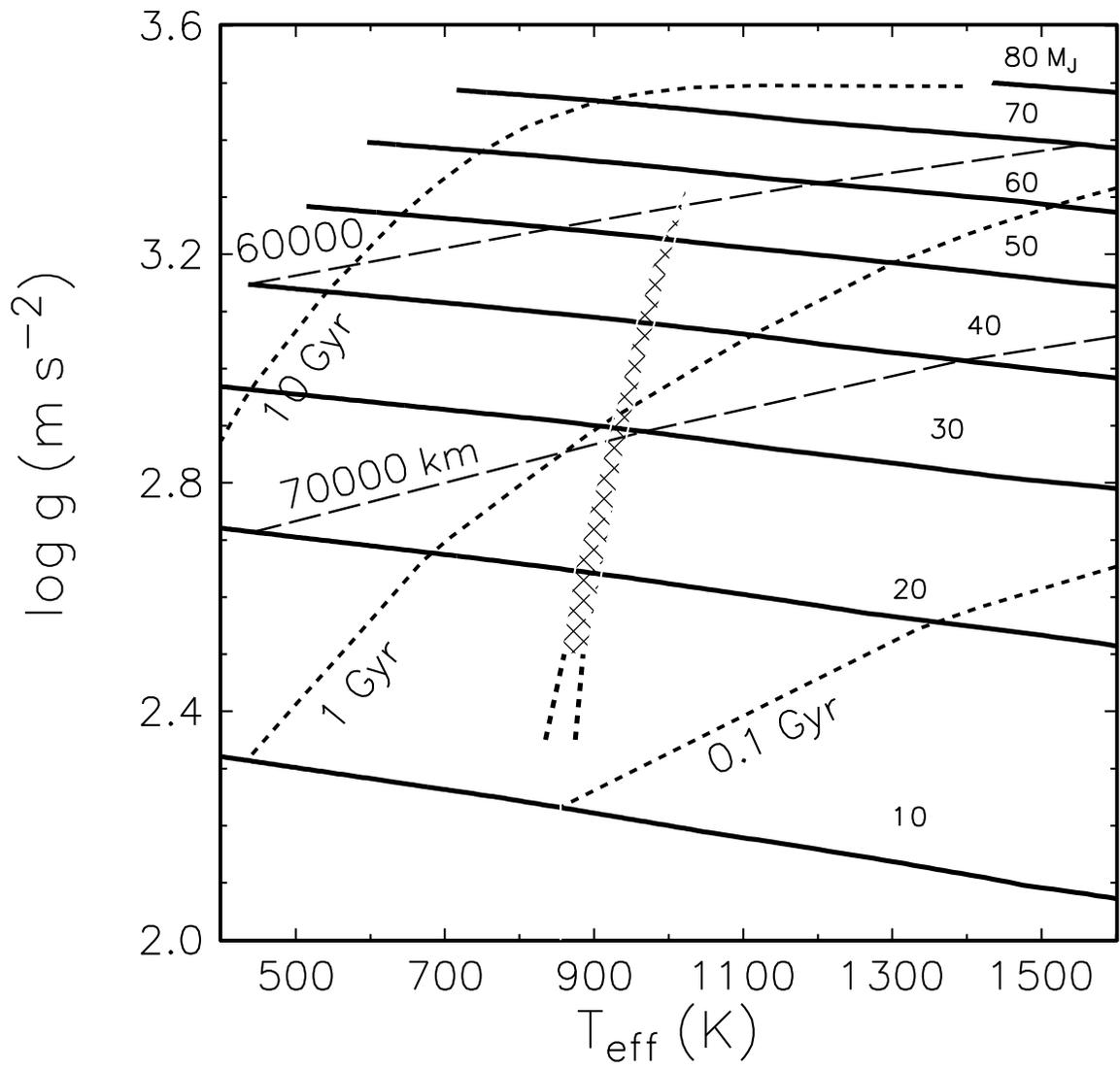
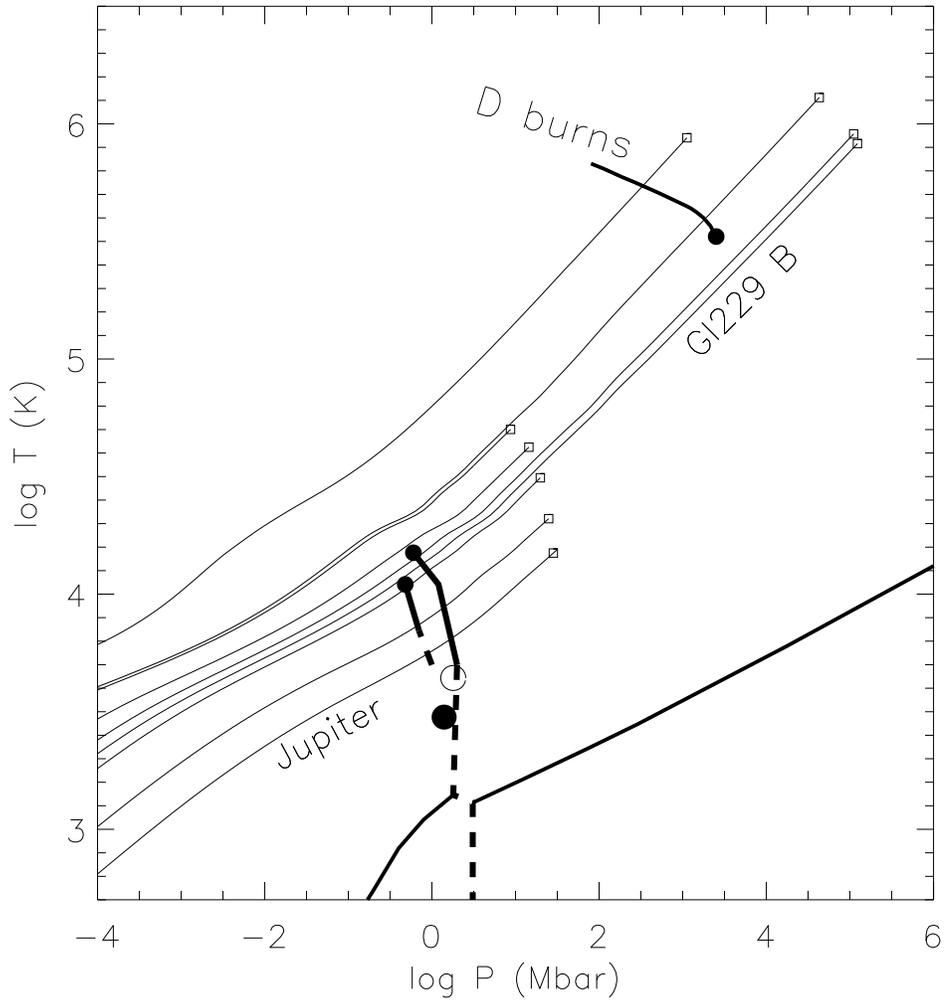
\end{document}